\documentclass[preprint,12pt]{elsarticle}

\usepackage{amssymb}
\usepackage{amsthm}
\usepackage{amsmath}
\usepackage{subfigure}

\journal{ }

\begin{document}

\begin{frontmatter}



\title{\bf Constrain the UT angle $\gamma$ by CP violation parameters in $B^0\to\pi^+\pi^-$}


\author[affiliation1]{Qin Qin}
\author[affiliation2]{Zhi-Tian Zou}
\author[affiliation1,affiliation4]{Cai-Dian L\"u}
\author[affiliation2]{Ying Li}

\address[affiliation1]{Institute of High Energy Physics, Beijing
100049, People's Republic of China}
\address[affiliation2]{Department of Physics, Yantai University, Yantai,
Shandong 264005, People's Republic of China}
\address[affiliation4]{State Key Laboratory of Theoretical Physics, Institute of Theoretical Physics,
Chinese Academy of Sciences, Beijing 100190, China}

\begin{abstract}
  We calculate the tree and penguin amplitudes in the $B^0\to\pi^+\pi^-$ decay channel
  employing the perturbative QCD factorization approach. Using the amplitudes as input
  with the theoretical uncertainties sufficiently considered, we constrain the UT
  angle $\gamma$ to $53^\circ\leq\gamma\leq70^\circ$, from the measurements of the CP
  violation parameters $C_{\pi^+\pi^-}$ and $S_{\pi^+\pi^-}$ in $B^0\to\pi^+\pi^-$.
  The U-spin breaking effect between $B^0\to\pi^+\pi^-$ and $B_s^0\to K^+K^-$ is
  estimated to be around 30\%.
\end{abstract}

\begin{keyword}
CP violation \sep $\gamma$ extraction \sep $B$ meson decay



\end{keyword}

\end{frontmatter}


\section{Introduction}

In the Standard Model, the quark mixing is described by the
Cabbibo-Kobayashi-Maskawa (CKM) matrix \cite{Kobayashi:1973fv}, in which the nonzero
phase angle induces the Charge conjugation Parity (CP) violation in weak interaction. For recent
developments on the CKM matrix, one can refer to the review \cite{Wang:2014sba}.
It is important to examine the unitarity of the CKM matrix, since any deviation
would indicate new physics beyond the Standard Model. The three angles of the well-known unitarity
triangle (UT), which are defined by $\alpha\equiv\arg[-(V_{td}V_{tb}^*)/(V_{ud}V_{ub}^*)]$,
$\beta\equiv\arg[-(V_{cd}V_{cb}^*)/(V_{td}V_{tb}^*)]$ and $\gamma\equiv\arg[-(V_{ud}V_{ub}^*)/ (V_{cd}V_{cb}^*)]$,
have been measured by experiments and the present averages are
\cite{Agashe:2014kda}
\begin{equation}\label{world_average}
\alpha = (85.4^{+3.9}_{-3.8})^\circ,~~~~\sin2\beta = 0.682\pm0.019,~~~~\gamma = (68.0^{+8.0}_{-8.5})^\circ.
\end{equation}
The angle $\gamma$ is the least known one among the three angles. Methods were proposed
to extract $\gamma$ from the tree-dominated modes $B\to DK$, known as the GLW method
\cite{Gronau:1990ra}, the ADS method \cite{Atwood:1996ci}, and
the Dalitz-plot method \cite{Giri:2003ty}, with different final states of $D$ decays.
Combining the $B\to DK$ measurements performed by Belle, BaBar, CDF and LHCb
\cite{Trabelsi:2013uj}, the CKMfitter group \cite{Charles:2011va} obtained the above
average for $\gamma$. Recently, the LHCb collaboration made two new measurements
\cite{Aaij:2014uva}. Alternatively, $\gamma$ can also be determined by the U-spin analysis
on the two-body charmless $B$ decays, $B^0\to\pi^+\pi^-$ and $B_s^0\to K^+K^-$
\cite{Fleischer:1999pa}. A combination with the channels $B^0\to\pi^0\pi^0$ and
$B^+\to\pi^+\pi^0$ makes the analysis more sophisticated \cite{Ciuchini:2012gd}. Recently,
following the method proposed in Ref. \cite{Ciuchini:2012gd}, the LHCb collaboration
performed the U-spin and isospin analysis and obtained \cite{Aaij:2014xba}
\begin{equation}
\gamma=(63.5^{+7.2}_{-6.7})^\circ,
\end{equation}
which has a smaller central value than the world average in Eq. (\ref{world_average}).

In this letter, we constrain the UT angle $\gamma$ from $B^0\to\pi^+\pi^-$, with the help
of factorization approach to calculate the tree and penguin amplitudes.
Similar ideas have been used to constrain $\alpha$ from $B^0\to\pi^+\pi^-$ \cite{Lu:2002qv},
and to constrain $\gamma$ from $B^0_s\to D_s^\pm K^\mp$ \cite{Yu:2013pua}. However, neither
of them got strong constraint on the corresponding UT angle for lack of precisely measured
experimental results at their time. Recently, the CP violation parameters in
$B^0\to\pi^+\pi^-$ have been precisely measured \cite{Lees:2012mma}, and the weighted
averages of the results are given by \cite{Aaij:2014xba},
\begin{equation}\label{CSexperiment}
C_{\pi^+\pi^-} = -0.30 \pm 0.05,~~~~S_{\pi^+\pi^-} = -0.66 \pm 0.06,
\end{equation}
with the statistical correlation $\rho(C_{\pi^+\pi^-},S_{\pi^+\pi^-})=-0.007$. The high
precision of the parameters indicates the possibility that our constraint on $\gamma$ is
comparable to the world average in Eq. (\ref{world_average}) and the
results given in Ref. \cite{Aaij:2014xba}. The method can also be applied to
$B_s^0\to K^+K^-$.

The rest of the paper is organized as follows. In Sec. \ref{formalism}, the relevant
formulas for the CP violation parameters in the channels $B^0\to\pi^+\pi^-$ and
$B_s^0\to K^+K^-$ are listed. In Sec. \ref{numeric}, we introduce our strategy for the
numerical analysis and obtain the constraints on $\gamma$ from the two channels, between
which the U-spin breaking effect is also estimated. In Sec. \ref{conclude}, we conclude.



\section{Theoretical formalism}\label{formalism}

For $B^0\to\pi^+\pi^-$, the relevant effective Hamiltonian is given by
\begin{equation}\begin{split}
\mathcal{H}_{eff}=&V_{ub}^*V_{ud}[C_1O_1+C_2O_2]
-V_{tb}^*V_{td}\sum_{n=3}^{10}C_nO_n+{\it h.c.},
\end{split}\end{equation}
where $O_{1,2(3-10)}$ are the tree (penguin) 4-quark operators, and $C_{1-10}$ are
the corresponding Wilson coefficients. After we apply some factorization approach
to calculate the hadronic matrix elements $\langle\pi^+\pi^-|O_i|B^0\rangle$,
the amplitude of $B^0\to\pi^+\pi^-$ can be expressed as
\begin{equation}\begin{split}
\mathcal{A}(B^0\to\pi^+\pi^-)=&V_{ub}^*V_{ud}\mathcal{T}-V_{tb}^*V_{td}\mathcal{P}\\
=&V_{ub}^*V_{ud}(\mathcal{T}+\mathcal{P})\left(1+{V_{cb}^*V_{cd}\over V_{ub}^*V_{ud}}{\mathcal{P}\over\mathcal{T}+\mathcal{P}}\right),
\end{split}\end{equation}
where $\mathcal{T}$ and $\mathcal{P}$ are the tree and penguin amplitudes, respectively.
Defining
\begin{equation}
de^{i\theta}\equiv{|V_{cb}^*V_{cd}|\over|V_{ub}^*V_{ud}|}{\mathcal{P}\over\mathcal{T}+\mathcal{P}},
\end{equation}
with $d$ and $\theta$ real-valued, we obtain the expression for the CP violation
parameters
\begin{equation}\begin{split}
C_{\pi^+\pi^-}=&-{2d\sin\theta\sin\gamma\over1+d^2-2d\cos\theta\cos\gamma},\\
S_{\pi^+\pi^-}=&-{\sin(2\beta+2\gamma)-2d\cos\theta\sin(2\beta+\gamma)+d^2\sin(2\beta)\over1+d^2-2d\cos\theta\cos\gamma}.
\end{split}\end{equation}
For $C_{\pi^+\pi^-}$ and $S_{\pi^+\pi^-}$, we have accepted the convention in the
letter \cite{Aaij:2014xba},
\begin{equation}\begin{split}
&C_{\pi^+\pi^-}\equiv{1-|\lambda_{\pi^+\pi^-}|^2\over1+|\lambda_{\pi^+\pi^-}|^2},~~~~
S_{\pi^+\pi^-}\equiv{2\mathrm{Im}\lambda_{\pi^+\pi^-}\over1+|\lambda_{\pi^+\pi^-}|^2},\\
&\lambda_{\pi^+\pi^-}\equiv{q\over p}{\mathcal{A}(\bar{B}^0\to\pi^+\pi^-)\over\mathcal{A}(B^0\to\pi^+\pi^-)},
\end{split}\end{equation}
where $q$ and $p$ are the coefficients in the mass eigenstates $p|B^0\rangle\pm q|\bar{B}^0\rangle$.

Similarly for $B_s^0\to K^+K^-$, one has
\begin{equation}\begin{split}\label{K2}
C_{K^+K^-}\approx&{2\widetilde{d'}\sin\theta'\sin\gamma\over1+\widetilde{d'}^2+2\widetilde{d'}\cos\theta'\cos\gamma},\\
S_{K^+K^-}\approx&-{\sin(-2\beta_s+2\gamma)+2\widetilde{d'}\cos\theta'\sin(-2\beta_s+\gamma)+\widetilde{d'}^2\sin(-2\beta_s) \over1+\widetilde{d'}^2+2\widetilde{d'}\cos\theta'\cos\gamma},
\end{split}\end{equation}
where the real-valued parameters are defined by
\begin{equation}\label{K3}
\widetilde{d'}\equiv{|V_{cs}||V_{ud}|\over|V_{cd}||V_{us}|}d',~~~~
d'e^{i\theta'}\equiv{|V_{cb}^*V_{cd}|\over|V_{ub}^*V_{ud}|}{\mathcal{P'}\over\mathcal{T'}+\mathcal{P'}},
\end{equation}
with $\mathcal{T'}$ ($\mathcal{P'}$) representing the tree (penguin) amplitude in
$B_s^0\to K^+K^-$. $\beta_s\equiv\arg[-(V_{ts}V_{tb}^*)/(V_{cs}V_{cb}^*)]$ gives the
mixing phase in the $B_s^0-\bar{B}_s^0$ mixing system.

\section{Numerical Analysis}\label{numeric}

\begin{figure}[!ht]
  \centering
  \subfigure[$C_{\pi^+\pi^-}$ {\it vs} $\gamma$;]{
    \includegraphics[width=2.4in]{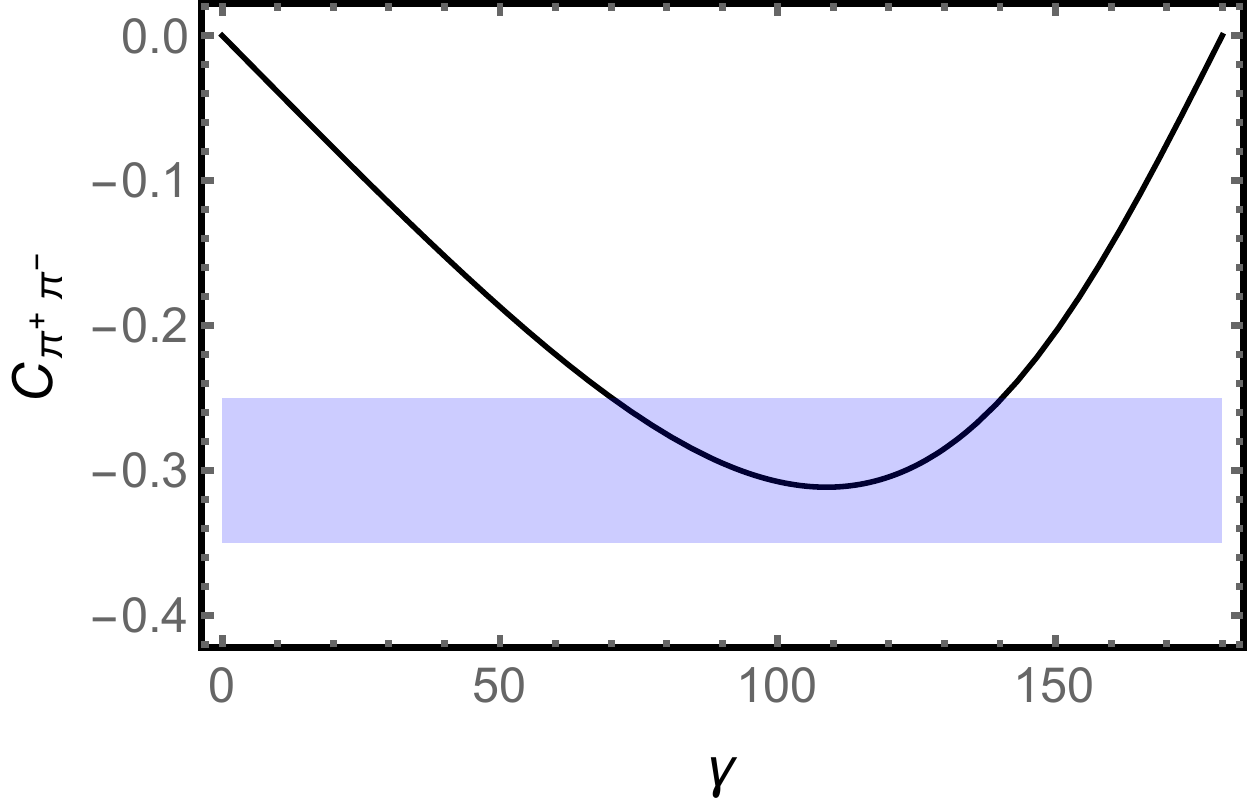}\label{Cpipi-Gamma}}
  \hspace{0.2in}
  \subfigure[$S_{\pi^+\pi^-}$ {\it vs} $\gamma$.]{
    \includegraphics[width=2.4in]{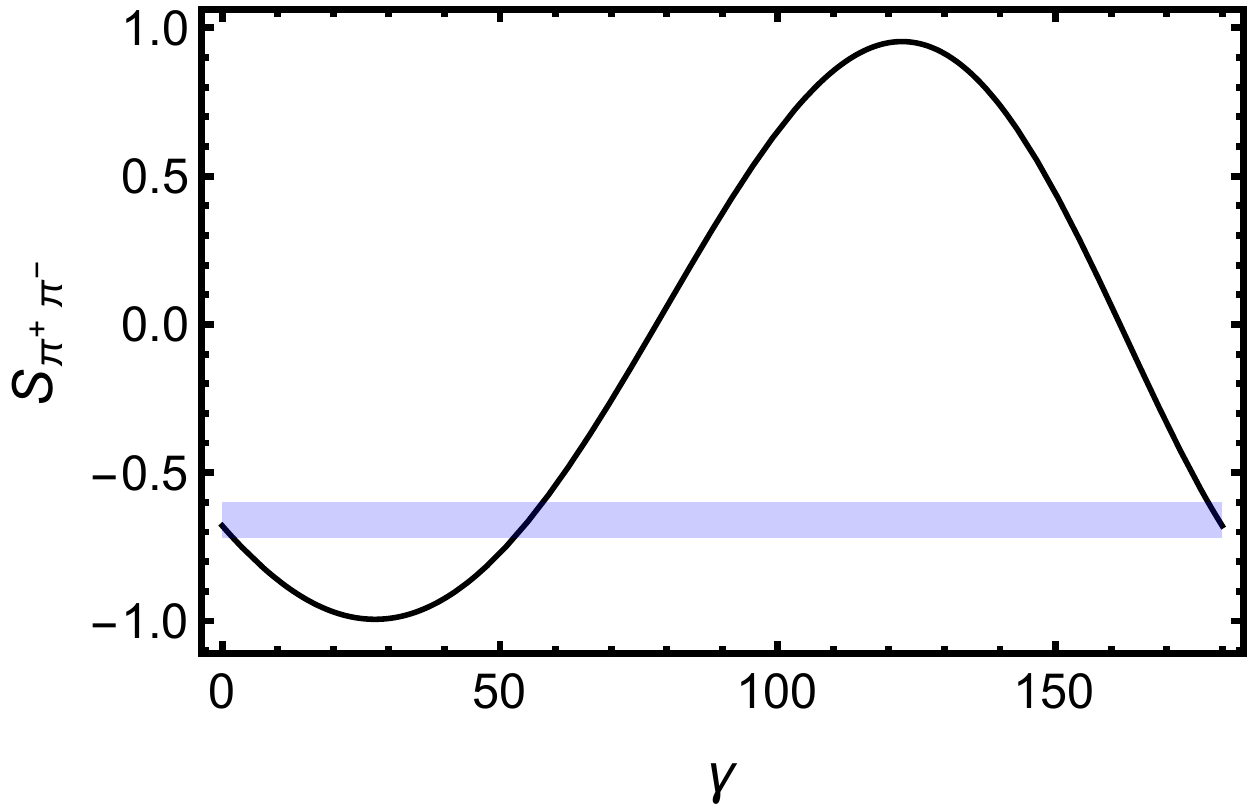}\label{Spipi-Gamma}}
  \caption{The solid curves correspond to the sample choice: $d=0.23$ and $\theta=2.4$.
  The light blue bands show the experimentally 1 $\sigma$ allowed regions
  $-0.35\leq C_{\pi^+\pi^-}\leq-0.25$ and $-0.72\leq S_{\pi^+\pi^-}\leq-0.60$, respectively.}
  \label{K_mixing_plot}
\end{figure}

The present average of the UT angle $\beta$ is given in Eq. (\ref{world_average}),
which has a two-fold ambiguity $2\beta\to\pi-2\beta$. A series of measurements
\cite{Aubert:2004cp} prefer that $\cos2\beta$ is positive, so we accept
\begin{equation}\label{beta_experiment}
\beta=(21.50_{-0.74}^{+0.75})^\circ.
\end{equation}
Choosing the sample values for $d$ and $\theta$, $de^{i\theta}=0.23e^{i2.4}$, we can
then obtain the $\gamma$ dependence of $C_{\pi^+\pi^-}$ and $S_{\pi^+\pi^-}$, as shown
in Fig. \ref{K_mixing_plot}. The experimental 1 $\sigma$ allowed regions are also
displayed. Fig. \ref{Spipi-Gamma} shows that $S_{\pi^+\pi^-}$ is very sensitive to the
change of the angle $\gamma$, and at meanwhile precise measurements for $S_{\pi^+\pi^-}$
have been performed. This indicates that $\gamma$ is potentially to be strongly constrained
in our method, though there are considerable theoretical uncertainties in any factorization
approach.

In the perturbative QCD (PQCD) approach base on the transverse momentum factorization
\cite{Chang:1996dw}, hadronic matrix elements are factorized into convolutions
of the calculable hard kernels and the non-perturbative meson wave functions which are
however universal. The PQCD approach has been applied in analysis on hadronic
$B$ meson decays, successfully making predictions for both branching ratios and
CP violation \cite{Lu:2000em,Ali:2007ff}. Especially for $B^0\to\pi^+\pi^-$, the
PQCD prediction of the branching ratio is
$(5.8^{+3.0+0.5+0.4}_{-2.1-0.4-0.3})\times10^{-6}$ \cite{Ali:2007ff}, which is
consistent with the experimental result
$(5.12\pm0.19)\times10^{-6}$ \cite{Agashe:2014kda}. Therefore, we employ the PQCD
approach to calculate the tree and penguin amplitudes here. The formulas for
calculating the leading-order decay amplitudes are given by Eqs. (50 - 61) in Ref.
\cite{Ali:2007ff}. We also include the next-to-leading-order corrections to the
$B\to\pi$ transition form factors, of which the twist-2 and -3 contributions have
been studied in Ref. \cite{Li:2012nk} and \cite{Cheng:2014fwa}, respectively.



\begin{figure}[!ht]
  \centering
    \includegraphics[width=2.5in]{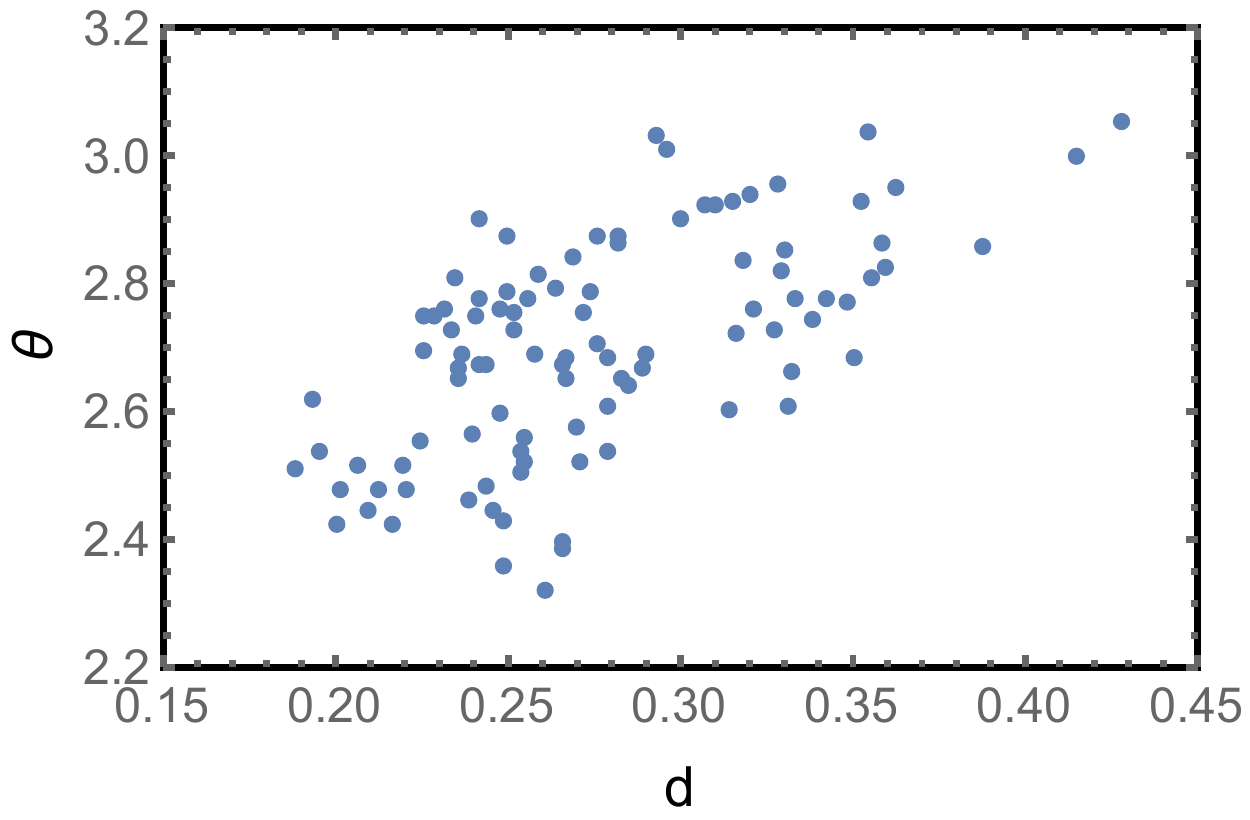}
  \caption{Plots for ($d$, $\theta$) calculated with the random theoretical parameters ranging in the allowed regions.}
  \label{d-theta} 
\end{figure}

\begin{figure}[!ht]
  \centering
    \includegraphics[width=2.5in]{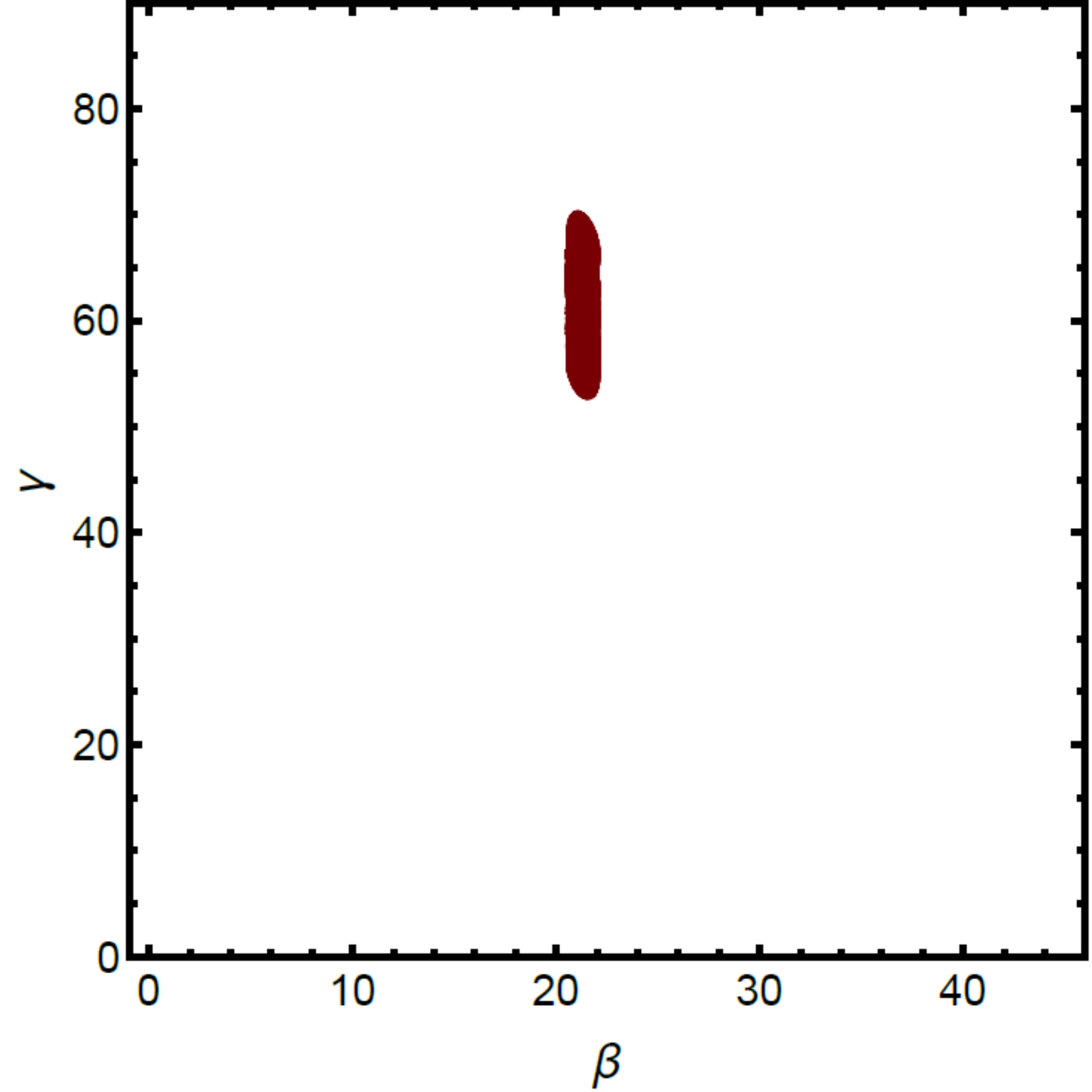}
  \caption{The 68\% C.L. region for $\beta$ - $\gamma$ with the theoretical uncertainties considered.}
  \label{fit_scan} 
\end{figure}

To perform a reliable analysis, we need to sufficiently take into account the
uncertainties introduced by the calculation of the hadronic matrix elements.
In the calculation, we adopt the updated non-asymptotic distribution amplitudes
\cite{Khodjamirian:2009ys},
\begin{equation}\begin{split}
\phi^A_\pi(x)=&{f_\pi\over2\sqrt{6}}6x(1-x)[1+a^\pi_2C_2^{3/2}(2x-1)+a^\pi_4C_4^{3/2}(2x-1)],\\
\phi^P_\pi(x)=&{f_\pi\over2\sqrt{6}}[1+30\eta_3^\pi C^{1/2}_2(2x-1)-3\eta_3^\pi\omega_3^\pi C_4^{1/2}(2x-1)],\\
\phi^T_\pi(x)=&{f_\pi\over2\sqrt{2N_c}}(1-2x)\{1
+{1\over2}\eta_3^\pi(10-\omega_3^\pi)C_2^{3/2}(2x-1)-15\eta_3^\pi(10-\omega_3^\pi)x(1-x)\},
\end{split}\end{equation}
where $C^{\alpha}_n(2x-1)$ are the well known Gegenbauer polynomials with $x$
the longitudinal momentum fraction of the quark in pion. The values of the Gegenbauer moments,
$a_2^\pi$ and $a_4^\pi$, have been determined in the global fit to the data of the
pion electromagnetic form factor \cite{Khodjamirian:2011ub}, which yields
\begin{equation}
a_2^\pi=0.17\pm0.08,~~~~a_4^\pi=0.06\pm0.10.
\end{equation}
To keep it safe, we double the error bars in the numerical analysis. In Ref.
\cite{Li:2013xna} where the joint resummation was performed for the pion transition
form factor in the transverse-momentum factorization formalism, the authors found
that their prediction for the form factor with
$a_2^\pi=0.05$ agrees well with the experimental data. Our choice for the range of
$a_2^\pi$ covers this value. As for the other non-perturbative parameters $\eta_3^\pi$
and $\omega_3^\pi$, we accept the values taken in Ref. \cite{Khodjamirian:2009ys}, also
with doubled error bars. The shape parameter in the distribution amplitude of the
$B^0$ meson \cite{Kurimoto:2001zj} is taken value in the range
\begin{equation}
\omega_b\in[0.36,~0.44].
\end{equation}
We also consider the uncertainties caused by the unknown next-to-leading-order corrections
characterized by the choice that $\Lambda_{QCD}\in[0.20,0.30]$ and a 20\% variation of the
factorization scale.
Taking values for the theoretical parameters randomly in the ranges covering
their uncertainties, we perform the PQCD calculation and obtain 99 points of
($d$, $\theta$), which are shown in Fig. \ref{d-theta}.
At each point of ($d$, $\theta$), we perform the global fit of $\beta$ and $\gamma$
to the experimental results of the CP violation parameters in Eq. (\ref{CSexperiment})
and that of $\beta$ in Eq. (\ref{beta_experiment}). Then, we combine the 1 $\sigma$
allowed regions of all fits at the 99 points, and regard it as our constraint on
$\gamma$ and $\beta$. As shown in Fig. \ref{fit_scan}, the constraint on $\gamma$ is
\begin{equation}\label{final_constraint}
53^\circ\leq\gamma\leq70^\circ.
\end{equation}


We can also perform the similar analysis to $B_s^0\to K^+K^-$ using Eqs. (\ref{K2}) and
(\ref{K3}), though the experimental results for the CP violation parameters are much
less precise, which are given by \cite{Aaij:2014xba}
\begin{equation}
C_{K^+K^-}=0.14\pm0.11,~~~~S_{K^+K^-}=0.30\pm0.13,
\end{equation}
with the statistical correlation $\rho(C_{K^+K^-},S_{K^+K^-})=0.02$.
To improve the precision on the determination of $\gamma$, $\beta_s$ is expressed in
terms of $\beta$ and $\gamma$. However, the $B_s^0\to K^+K^-$ constraint
$20^\circ\leq\gamma\leq150^\circ$ is still too loose.

As a byproduct, we also estimate the U-spin breaking effect in the two channels
$B^0\to\pi^+\pi^-$ and $B_s^0\to K^+K^-$, which is parameterized by
\begin{equation}
d'e^{i\theta'}=de^{i\theta}(1+re^{i\theta_r}).
\end{equation}
The PQCD result is
\begin{equation}
r=0.3\pm0.1,~~~~\theta_r=-1.2\pm0.2.
\end{equation}
In the letter \cite{Aaij:2014xba}, the U-spin breaking effect is parameterized
by two relative magnitudes $r_D$ and $r_G$ with the corresponding phase shifts
$\theta_{r_D}$ and $\theta_{r_G}$,
\begin{equation}
d'e^{i\theta'}=de^{i\theta}{1+r_Ge^{i\theta_{r_G}}\over1+r_De^{i\theta_{r_D}}}.
\end{equation}
Assuming the parameters range within the region
\begin{equation}
r_D,r_G\in[0,~0.5],~~~~\theta_{r_D},\theta_{r_G}\in[-\pi,~\pi],
\end{equation}
the authors obtained $\gamma=(63.5^{+7.2}_{-6.7})^\circ$.
This region can fully cover the PQCD result (including the uncertainties),
so we conclude that the assumption about the U-spin breaking in Ref.
\cite{Aaij:2014xba} is reasonable.

\section{Conclusion}\label{conclude}

We extract the UT angle $\gamma$ from the precise experimental results of
$C_{\pi^+\pi^-}$ and $S_{\pi^+\pi^-}$ given in the letter \cite{Aaij:2014xba},
with the tree and penguin amplitudes in $B^0\to\pi^+\pi^-$ calculated in the PQCD
approach. Including the theoretical uncertainties, we constrain $53^\circ\leq\gamma\leq70^\circ$
at 68\% probability. Through the similar method, the angle $\gamma$ is also
constrained in the range $20^\circ - 150^\circ$ by the measurements of $C_{K^+K^-}$ and
$S_{K^+K^-}$. The U-spin breaking effect between the two channels is found to be
smaller than 50\%, which indicates that the results in the letter \cite{Aaij:2014xba}
are reliable.

\section*{Acknowledgement}

We are grateful to Shan Cheng, Wei Wang and Yu-Ming Wang for useful discussions.
QQ thanks Yantai University for the warm hospitality during his visit. This work
is partly supported by the National Science Foundation of China under Grants No.
11375208, No. 11235005 and No. 11447032, the Natural Science Foundation of Shandong
province (ZR2014AQ013) and the Program for New Century Excellent Talents in
University (NCET) by Ministry of Education of P. R. China (Grant No. NCET-13-0991).





\end{document}